\documentclass[aps,prl,longbibliography,twocolumn]{revtex4-2}
\pdfoutput=1 
\usepackage{CJK}

\usepackage{graphicx}

\usepackage{amsmath}
\usepackage{dcolumn}

\usepackage{bm}
\usepackage[normalem]{ulem}
\usepackage{color}
\usepackage{hyperref}

\usepackage{epstopdf}
\newcommand{\bew}{\begin{widetext}}
\newcommand{\ew}{\end{widetext}}

\newcommand{\tri}{\triangle}

\newcommand{\sep}{ \ \ \ , \ \ \ }

\newcommand{\beq}{\begin{equation}}
\newcommand{\eeq}{\end{equation}}
\newcommand{\beqn}{\begin{eqnarray}}
\newcommand{\eeqn}{\end{eqnarray}}
\newcommand{\pp}{\partial}
\newcommand{\dd}{{\rm d}}

\newcommand{\la}{\langle}

\newcommand{\ra}{\rangle}

\begin{document}
\title{Nonequilibrium glass transitions in the spherical $p$-spin model with antisymmetric interactions}
\author{Chiu Fan Lee}
\email{c.lee@imperial.ac.uk}
\address{Department of Bioengineering, Imperial College London, South Kensington Campus, London SW7 2AZ, U.K.}

\begin{abstract}
Our theoretical understanding of glassy dynamics is notoriously incomplete, and it is even more so when the glassy systems are driven out of equilibrium. 
An extreme way to drive a system out of equilibrium is to introduce nonequilibrium dynamics at the microscopic level, e.g., through active forcing  of the constituent particles or by having nonreciprocal interactions among the particles. While glassy dynamics under active forcing has been studied by many, the latter nonequilibrium scenario has received little attention. Here,  I study the glassy dynamics of the spherical $p$-spin model for $p\geq 3$ 
with antisymmetric interactions, which generalizes reciprocal interactions in 2-body interactions. 
The spherical $p$-spin model is an integral tool in the study of dynamical glass transition, and 
when antisymmetric interactions are added, I show analytically and numerically that glassy behavior is generically suppressed. Moreover, I obtain analytical expressions on the modified dynamical glass transition point and the Edward-Anderson parameter (i.e., the asymptotic plateau height value of the spin-spin correlation function) in the small driving limit.
	\end{abstract}
	
\maketitle

\begin{figure}
		\begin{center}
		\includegraphics[scale=.5]{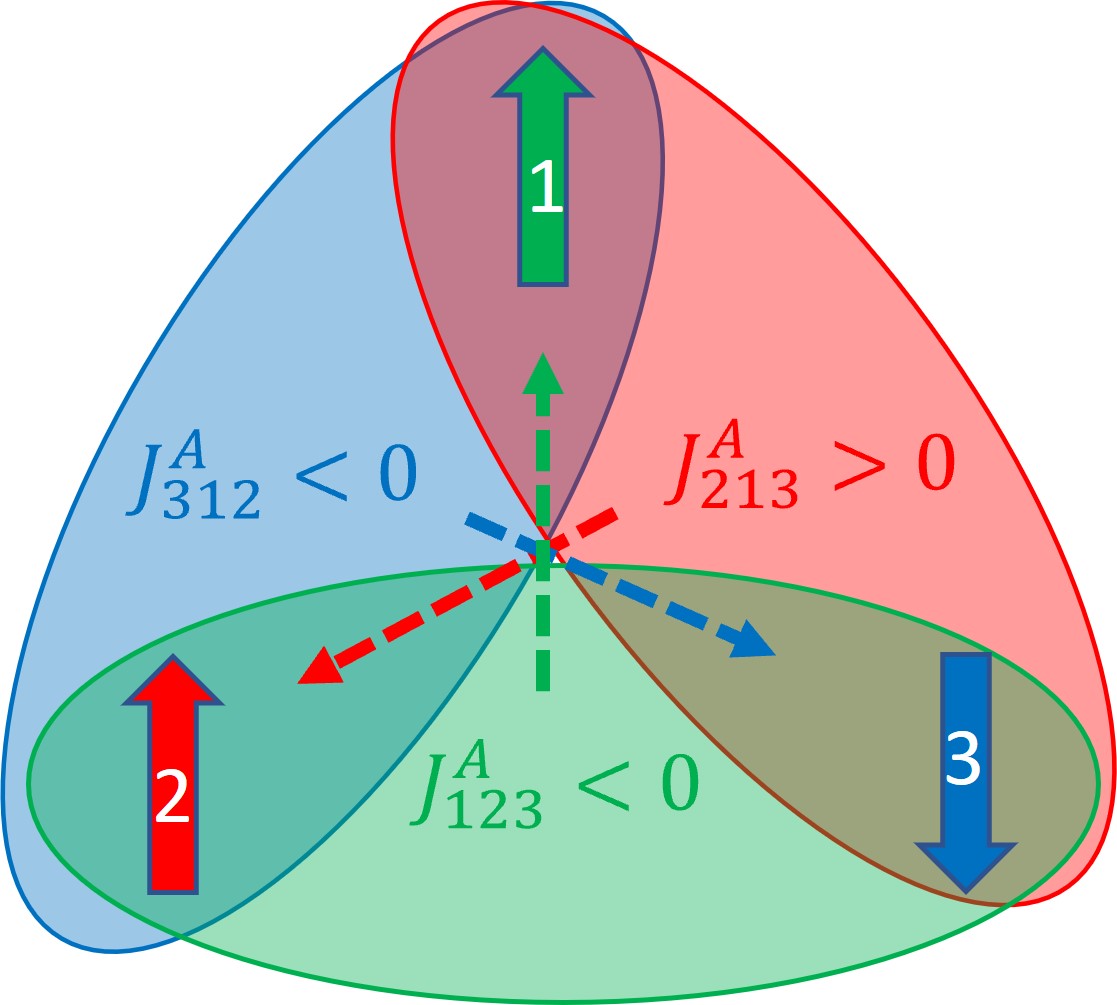}
		\end{center}
		\caption{{\it A schematic of a 3-spin system with  antisymmetric interactions.}		
		The antisymmetric 3-body interactions shown generalizes the 2-body nonreciprocal  interactions. Here, the force felt by spin $i$ is given by $J^A_{ijk}  \sigma_j \sigma_k $ where $j<k$ and $\sigma$'s denote the values of the spins. Since the interaction tensor $J^A$ is antisymmetric, these force terms cannot be obtained from differentiating a Hamiltonian, the system is therefore necessary nonequilibrium.
		} 
		\label{fig:cartoon}
	\end{figure}

The liquid-glass transition is a common occurrence in a many-body system as its density increases. Given the ubiquitous nature of glassy behavior, it is surprising that a comprehensive theoretical underpinning is still lacking \cite{
parisi_rmp10,berthier_rmp11,parisi_b20}. This is not only true for thermalizing systems, but even more so for driven, nonequilibrium systems.  
Glassy behavior has a wide spectrum of characteristics, ranging from the caging phenomenon at the molecular level to the macroscopic rheological properties that emerge as liquids turn into glasses. One key signature of glass transitions is the appearance of a plateau in the temporal decay of the intermediate scattering function \cite{reichman_jsm05,castellani_jsm05,hansen_b13}, 
which probes the local movement of particles. The liquid physics-based mode-coupling theory  provided the first dynamical description of this emergent glassy phenomenon \cite{leutheusser_pra84,bengtzelius_jpc84,das_rmp04}. Intriguingly, the resulting dynamical equation can be mapped onto the equation of motion (EOM) of the spherical  $p$-spin model when $p=3$ \cite{kirkpatrick_prl87,kirkpatrick_prb87}. The $p$-spin model itself is a highly schematic model of spin glasses, where the Hamiltonian is a sum of all combinations of $p$-tuples of spins with prefactors drawn from a certain probability distribution. 
Since then, the mode-coupling theory and the $p$-spin model have become  integral tools in the study of glassy dynamics for thermalizing systems \cite{crisanti_zpb93,cugliandolo_prl93}. These theoretical constructs have also recently been applied to driven nonequilibrium systems, such as systems of self-propelled particles in the high-density limits \cite{kranz_prl10,szamel_pre15,nandi_softmatt17,feng_softmatt17,janssen_frontier18}, and $p$-spin systems with active forcing \cite{berthier_natphys13}. 
Violating the fluctuation-dissipation relatiom at the microscopic level due to particles' self-generated forces, active matter is an extreme type of nonequilibrium system, and has proven to be a fertile ground of novel physics \cite{toner_ann05,ramaswamy_annrev10,marchetti_rmp13,bechinger_rmp16,chate_annrev20}. However, active forcing is not the only way to drive a system far from equilibrium at the microscopic level--nonreciprocal interactions among the constituents  is another way to do so and has started to attract increasing attention in the physics community \cite{ivlev_prx15,you_pnas20,saha_prx20,dadhichi_pre20,fruchart_nature21,loos_prl23}. 
Here, I study the glassy dynamics of $p$-spin model (for $p\geq 3$) with antisymmetric interactions, which generalize reciprocal interactions in 2-body interactions (Fig.~\ref{fig:cartoon}).
By studying analytically and numerically the dynamics of these $p$-spin systems, I show that glassy behavior is generically suppressed by the nonequilibrium, antisymmetric interactions. I further quantify their impact on the dynamical glass transition point, and the Edward-Anderson parameter (i.e., the asymptotic plateau height value of the spin-spin correlation function) under the {\it weak ergodicity breaking} approximation scheme \cite{bouchaud_jdp92,cugliandolo_prl93,cugliandolo_a02}.

{\it Antisymmetric $3$-spin model.---}While all analytical treatments in this Letter are on the general spherical $p$-spin model for  $p\geq 3$, I will discuss exclusively the $3$-spin model here from now on to simplify notation  and leave the more general case to the Supplemental Material \cite{SM}.

	\begin{figure}[t]
		\begin{center}
		\includegraphics[scale=.57]{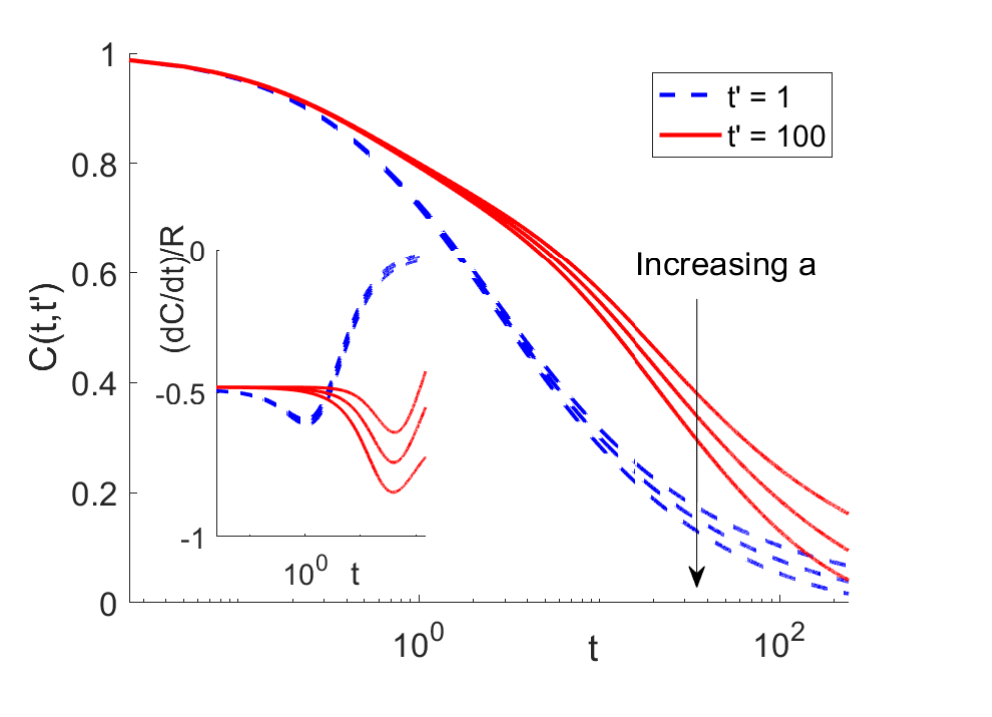}
		\end{center}
		\caption{{\it Temporal evolutions of the two-time correlation function and the fluctuation-dissipation relation (FDR).} Despite the presence of antisymmetric interactions ($a = 0, 0.08, 0.12$), the two-time correlation functions $C(t,t')$ continues to show the characteristic glassy behavior at $T=0.5$ below the dynamic glass transition noise strength $T_d$ (Eq.~\ref{eq:Td}). These characteristics include the appearance of a shoulder as $t'$ increases (red curves), and the applicability of the FDR ($T = -R^{-1} \pp_t C$) during the initial decay of $C$, even with the same noise strength $T=0.5$ irrespective of what $a$ is (inset figure.) 
The numerical algorithm used to generate these curves is identical to the one described in \cite{saraomannelli_prx20,ros_scipost21}, with the time increment $\tri t$ set to 0.025.		
		} 
		\label{fig:fig2}
	\end{figure}

The EOM of the spherical $3$-spin model consisting of $N$ spins is 
\beq
\label{eq:sigma0}
\frac{\dd \sigma_i(t)}{\dd t} = \sum_{j<k; j,k\neq i} J_{ijk} \sigma_{j} \sigma_{k} -\mu(t) \sigma_i(t) +\eta_i(t)
\eeq
where $i=1, \ldots, N$, $\eta_i$ is Gaussian noise term such that
\beq
\la \eta_i(t) \ra=0 \sep 
\la \eta_i(t)\eta_j(t') \ra=2T \delta_{ij}\delta(t-t') 
\ ,
\eeq
with $T$ being the noise strength (equivalent to the {\it temperature} in thermal systems),
and $\mu(t)$ is a time-dependent Lagrange multiplier to ensure that the average spin length in the system is one. Finally, the coupling terms $J_{ijk}$ are random variables drawn from a certain distribution. For thermal system, since the interaction terms for the $i$-th spin are obtained as follows:
\beq
\sum_{j<k; j,k\neq i} J^{\rm eq}_{ijk} \sigma_{j} \sigma_{k} 
= -\frac{\pp H[\sigma]}{\pp \sigma_i}
\ ,
\eeq
for some Hamiltonian $H$, $J^{\rm eq}_{ijk}$ are necessarily symmetric rank-3 tensors, i.e., $J^{\rm eq}_{ijk}$ are identical under permutations of the indices. 

To generalize beyond this equilibrium model, I will assume that
 the coupling coefficients $J$ is of the form
\beq
\label{eq:def_J}
J_{ijk} = J^{S}_{ijk}+aJ^{A}_{ijk}
\ ,
\eeq
where $a$ is a constant, and $J^S$'s and $J^A$'s are symmetric and antisymmetric tensors, respectively. Specifically, the sign of $J^{A}_{ijk}$ alternates when any two indices are interchanged. In the case of the $2$-spin model, it corresponds exactly to two-body nonreciprocal interactions and has already been studied in the context of neural networks \cite{crisanti_pra87}. However, the glassy dynamics that I am after is only manifested in $p$-spin models with $p\geq 3$, hence my focus here. Following Ref.~\cite{crisanti_pra87}, I assume that the entries in $J^S$'s and $J^A$'s are random variables drawn from a Gaussian distribution with zero means and variance $3/[N^2(1+a^2)]$. The choice of the assigned variance is to ensure that the variance of the tensor $J$ stays as $3/2N^2$, irrespective of whether $a$ is zero or not to facilitate comparison between the thermal and the nonequilibrium cases.

In the limit of $N$ going to infinity in this fully connected spin model, the EOM of the spin dynamics, $\sigma(t)$, with the quenched disorder averaged out  can be obtained from applying the saddle-point method on the corresponding Martin-Siggia-Rose-Janssen-De Dominicis generating functional \cite{SM}:
\beq
\label{eq:sigma1}
\frac{\dd \sigma}{\dd t}=-\mu \sigma + 3\alpha \int \dd t' R(t,t')C(t,t')^2\sigma(t')+\xi
\ ,
\eeq
where 
\beq
\alpha = \frac{1-2a^2}{1+a^2}
\ ,
\eeq
$\xi(t)$ is a noise term with statistics: 
\beq
\la \xi(t)\xi(t')\ra =2 T \delta (t-t') + \frac{3}{2} C(t,t')^{2}
\ ,
\eeq
and $C(t,t')$ and $R(t,t')$ are the two-time correlation and response functions, respectively. 
Specifically, $C(t,t') = \la \sigma(t) \sigma(t') \ra$ and $R(t,t') = \pp \la \sigma(t) \ra / \pp h(t')$ where $h(t')$ is an infinitesimal  external field that is turned on around time $t'$. Due to causality, $R(t,t')=0$ if $t<t'$. From the EOM, it is clear that since antisymmetric interactions generically decrease $\alpha$, which controls the positive,  time-integrated feedback on the spin dynamics, one may expect the correlation function would decay faster as the level of antisymmetry goes up. As we will see later, this is indeed the case from the numerical results.

From the EOM of $\sigma$, one can further derive the following closed set of couple equations of $C$ and $R$ that can be readily solved numerically:
\begin{widetext} 
\beqn
\label{eq:C}
\frac{\pp C(t,t')}{\pp t} &=& -\mu(t) C(t,t') + 3 \alpha\int_{0}^t  \dd t'' R(t,t'')C(t,t'')C(t'',t')
 +2TR(t',t)+\frac{3}{2} \int_{0}^{t'}  \dd t'' R(t',t'')C(t,t'')^2
\\
\label{eq:R}
\frac{\pp R(t,t')}{\pp t} &=& -\mu(t) R(t,t') + 3\alpha \int_{t'}^t  \dd t'' R(t,t'')C(t,t'')R(t'',t') +\delta(t,t')
\\
\label{eq:mu}
\mu(t) &=& T+\frac{3(1+2\alpha)}{2} \int_{0}^{t}  \dd t'' R(t,t'')C(t,t'')^2
 \ .
\eeqn
\end{widetext} 
From the form of the above EOM, one can now conclude that the effect of antisymmetric interactions {\it cannot} simply be accounted for by an effective noise strength $T_{eff}$ within the $p$-spin model.

For the thermal 3-spin model, it is known that there is a dynamic glass transition at $T_d= \sqrt{3/8} \simeq 0.61$. For $T$ above $T_d$, $C(t,t')$ decays to zero in $t$, irrespective $t'$, during which time the fluctuation-dissipation relation (FDR): $R = T^{-1} \pp C/\pp t$, is also satisfied. Below $T_d$, a shoulder develops in the decay of $C$ as  $t'$ grows, which becomes a plateau as $t' \rightarrow \infty$. However, the FDR seems to still remain valid during the initial decay of $C$, approximately until $t$ reaches the shoulder \cite{barrat_a97,cugliandolo_a02,castellani_jsm05}. 

Interestingly, the same phenomenology seems to still apply when antisymmetric interactions are introduced, at least in the small driving ($a\ll 1$) regime, as shown in Fig.~\ref{fig:fig2} obtained by numerical integrating the EOM (\ref{eq:C}--\ref{eq:mu}). Intriguingly, the FDR seems to remain satisfied
in the initial phase of the decay in $C$, {\it even with the same noise strength $T$ as in the symmetric case} (see the inset of Fig.~\ref{fig:fig2}). These observations motivate 
the adoption of approximation schemes below, which are typically used in thermal $p$-spin to enable analytical progress.

{\it Glass transition.---}In thermal $p$-spin model at high $T$, the relaxation follows equilibrium dynamics such that $C(t,t') = C_{\rm eq}(t-t')$ and $R(t,t') = R_{\rm eq}(t-t')$, and the FDR applies. In particular, one could use the FDR to eliminate the response field in the coupled EOM (\ref{eq:C}--\ref{eq:mu}), leading to the following reduced EOM
\begin{align}
\nonumber
&\frac{\pp C_{\rm eq}(t-t')}{\pp t} = -\mu(t) C_{\rm eq}(t-t')\\
&
\nonumber
\ \ \
 - \frac{3\alpha}{T} \int_{-\infty}^t  \dd t'' C_{\rm eq}'(t-t'')C_{\rm eq}(t-t'')C_{\rm eq}(t''-t')
\\
& 
\ \ \ -\frac{p}{2T} \int_{-\infty}^{t'}  \dd t'' C_{\rm eq}'(t'-t'')C_{\rm eq}(t-t'')^2
\\
&\mu(t) = T-\frac{6\alpha +3}{2T} \int_{-\infty}^{t}  \dd t''  C_{\rm eq}'(t-t'')C_{\rm eq}(t-t'')^2
\ ,
\end{align}
where the thermal case is recovered by setting $\alpha$ to 1.

	\begin{figure}[t]
		\begin{center}
		\includegraphics[scale=.5]{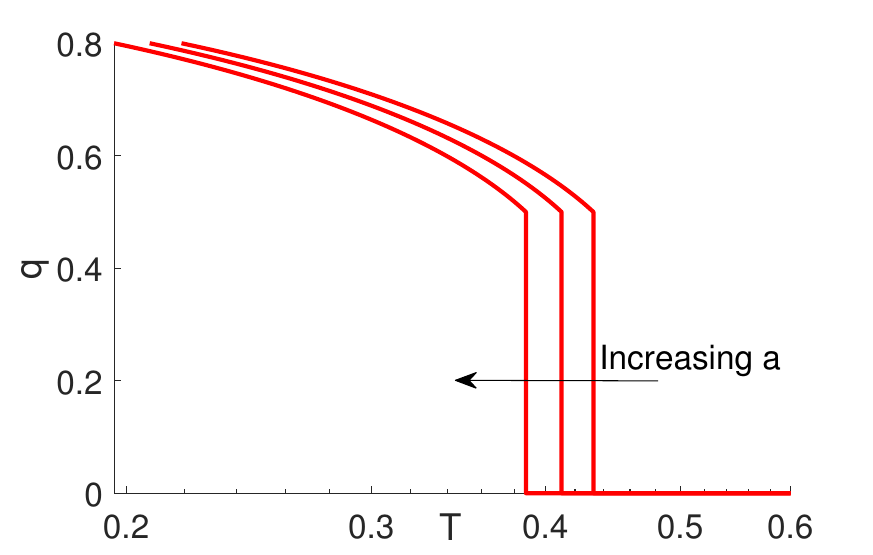}
		\end{center}
		\caption{{\it Asymptotic plateau height value $q$ vs.~noise strength $T$.}		
		 Increasing the antisymmetric contribution to the interactions ($a=0, 0.08, 0.12$) leads generically to 1) the decreases of the dynamic glass transition noise threshold, $T_d$ (\ref{eq:Td}), at which the plateau height value jumps from 0 to $1/2$; and 2) the decrease of the plateau height value $q$ (\ref{eq:q}). 
		} 
		\label{fig:plateau}
	\end{figure}
	
 Now, assuming that $\lim_{\tau \rightarrow \infty} C_{\rm eq}(\tau) =q$ where $q$ is the plateau value, also termed the {\it Edward-Anderson} parameter.  A dynamical glass transition  occurs at the highest $T$ value at which a nonzero $q$ value can be obtained self-consistently from the EOM. Solving the steady state equation with the required asymptotic restriction using the standard procedure \cite{SM}, we arrive at the following dynamic glass transition noise strength $T_d$: 
 \beq
 \label{eq:Td}
 T_d = \sqrt{\frac{\alpha-1}{2} +\frac{3}{8}}
 \ ,
 \eeq
 with the EA parameter $q$ being 1/2. Namely, the plateau height value remains unchanged from the symmetric case even  with the addition of antisymmetric interactions (the same conclusion applies for general $p$ \cite{SM}.)

{\it Aging regime.---}Below $T_d$, the assumption on the equilibrium-like dynamics is no longer valid, and we enter the so-called {\it aging regime}, in which the system's dynamics depend crucially on both $t$ and $t'$. To proceed analytically, I will adopt the following assumptions under the so-called {\it weak ergodicity breacking scenario} \cite{bouchaud_jdp92,cugliandolo_prl93,cugliandolo_a02}, which are typcially used in thermalizing $p$-spin models:
 \beqn
 \label{eq:formC}
 C(t,t')&=& C_{\rm eq}(t-t') + C_{\rm ag} (t,t')
 \\ 
 \label{eq:formR}
 R(t,t')&=& R_{\rm eq}(t-t') + R_{\rm ag} (t,t')
 \ ,
 \eeqn 
where $C_{\rm eq}(\tau)$ and $R_{\rm eq}(\tau)$ account for the initial {\it equilibrium-like} dynamics, while $C_{\rm ag}(t,t')$ and $R_{\rm ag}(t,t')$ account for the dynamics in the {\it aging} regime. Specifically, we now assume that $C_{\rm eq}(0) = 1-q$, $R_{\rm eq} (\tau) = -\frac{1}{T} C_{\rm eq}'(\tau) \Theta(\tau)$, and $C_{\rm ag} (0,0)=q$, where $q$ is now the plateau-height value of $C$ in limits: $\lim_{t-t' \rightarrow \infty}\lim_{t' \rightarrow \infty} C(t,t')$. With the approximated forms specified in Eqs~(\ref{eq:formC}) \& (\ref{eq:formR}), one can again follow the standard procedure in thermal $p$-spin model to find the relation between the plateau value $q$ and the noise strength $T$ for $T<T_d$ \cite{SM}:
 \beq
 \label{eq:q}
 T = \sqrt{\frac{3\alpha q}{2}}(1-q)
 \ .
 \eeq
 
Combining Eqs (\ref{eq:Td}) \& (\ref{eq:q}), Fig.~\ref{fig:plateau} shows how increasing the antisymmetric contribution to the interactions (\ref{eq:def_J}) generically lowers the glass transition noise threshold, $T_d$, and lowers the plateau height value $q$. This may be seen to be  consistent with the expectation that antisymmetric interactions are intrinsically nonequilibrium, whose presence constitutes an additional forcing in the system that promotes melting of the glass phase.

{\it Summary \& Outlook.---}I analyzed analytically and numerically the dynamical $p$-spin models ($p\geq 3$) with antisymmetric interactions, which are $p$-body generalizations of nonreciprocal interactions for $p=2$. The key results are 1) 
the finding that antisymmetric interactions generically suppress glassy behavior, 2) such a suppression can {\it not} be accounted for by an effective noise strength, and 3) the analytical expressions for the glass transition noise strength $T_d$ and the Edward-Anderson plateau height value $q$ as a function of the antisymmetry parameter. Interesting future directions include the generalization of the current investigation to the mixed $p$-spin models, which shows intricate memory effects \cite{folena_prx20}, and the study of the interplay between nonequilibrium antisymmetric interactions and the simultaneous presence of active forces \cite{berthier_natphys13}.

%

\end{document}


\title{Supplemental Material:\\
Nonequilibrium glass transitions in the spherical $p$-spin model with antisymmetric interactions}
\author{Chiu Fan Lee}
\email{c.lee@imperial.ac.uk}
\address{Department of Bioengineering, Imperial College London, South Kensington Campus, London SW7 2AZ, U.K.}

\maketitle

\section{Equation of Motion (EOM) of the correlation and response functions}
The EOM of the $p$-spin model  is
\beq
\pp_t \sigma_i(t) =  \sum_{i_1<i_2< \cdots < i_{p-1}} J_{ii_1 \cdots i_{p-1}} \sigma_{i_1} \cdots  \sigma_{i_{p-1}}  -\mu  \sigma_i+\xi_i
\ ,
\eeq
where $i=1, \ldots, N$, $\eta_i$ is Gaussian noise term such that
\beq
\la \eta_i(t) \ra=0 \sep 
\la \eta_i(t)\eta_j(t') \ra=2T \delta_{ij}\delta(t-t') 
\ ,
\eeq
and $\mu(t)$ is a time-dependent Lagrange multiplier to ensure that the average spin length in the system is one.
The coupling coefficients are of the form 
\beq
J_{i_1 \cdots i_p} = J_{i_1 \cdots i_p}^S + a J_{i_1 \cdots i_p}^A \ ,
\eeq
where $J^S$'s are completely symmetric tensor (but with vanishing entries that are associated with repeated indices) and $J^A$'s are anti-symmetric tensors. Besides these constraints, the entries of these tensors are drawn from a Gaussian distribution with zero mean and variance, i.e., the entries are drawn from the distribution \cite{s_crisanti_pra87}:
\beq
P(x)\propto \exp\left[\frac{2N^{p-1} (1+a^2)x^2}{p!} \right]\ .
\eeq
The choice of the assigned variance is to ensure that the variance of the tensor $J$ stays as
\beq
\frac{p!}{2N^{p-1}} \ ,
\eeq
irrespective of whether $a$ is zero or not.

Using the Martin-Siggia-Rose-Janssen-De Dominicis generating functional formalism, we arrive at \cite{s_kirkpatrick_prl87,s_kirkpatrick_prb87}
\beq
Z[J,h,\hat{h}] = \int \cD\sigma \cD \hat{\sigma}
 \exp \left\{
\int \dd t \left[ h_i \sigma_i + \ii \hat{h}_i \hat{\sigma}_i \right] + L(\sigma, \hat{\sigma})\right\}
\eeq
where
\beq
L(\sigma, \hat{\sigma}) 
=\int \dd t \sum_i \ii \hat{\sigma}_i \left[
\pp_t \sigma_i +\mu \sigma_i-\sum_{i_1<i_2< \cdots < i_{p-1}} J_{ii_1 \cdots i_{p-1}} \sigma_{i_1} \cdots  \sigma_{i_{p-1}} + D \ii \hat{\sigma}_i\right]
\ .
\eeq
We now focus on the following integral to integrate out the $J$'s:
\beqn
\nonumber
&&\int \cD [J^s] \cD [J^a] P([J^s]) P([J^a])\exp\bigg\{
-\int \dd t\sum_{\vec{n}_p}\bigg[ J_{\vec{n}_p}^s \left[
(\ii \hat{\sigma}_{i_1} \sigma_{i_2} \cdots  \sigma_{i_{p}}) + (\sigma_{i_1}\ii \hat{\sigma}_{i_2} \cdots  \sigma_{i_{p}})\cdots +
(\sigma_{i_1} \cdots  \sigma_{i_{p-1}}\ii \hat{\sigma}_p) 
\right]
\\
&&
+J_{\vec{n}_p}^a \left[
(\ii \hat{\sigma}_{i_1} \sigma_{i_2} \cdots  \sigma_{i_{p}}) - (\sigma_{i_1}\ii \hat{\sigma}_{i_2} \cdots  \sigma_{i_{p}})+ (\sigma_{i_1}\sigma_{i_2}\ii \hat{\sigma}_{i_3} \cdots  \sigma_{i_{p}})-
\cdots (\pm 1)^{p+1}
(\sigma_{i_1} \cdots  \sigma_{i_{p-1}}\ii \hat{\sigma}_p) 
\right]
 \bigg]
  \bigg\}
  \\
  &=& 
  \label{eq:S_int}
\exp\left\{
\int \frac{\dd t \dd t'}{4N^{p-1}} \left[ 
p(\ii \hat{\sigma} \cdot  \ii \hat{\sigma}) (\sigma \cdot \sigma )^{p-1}+
p(p-1)\alpha (\ii \hat{\sigma} \cdot  {\sigma}) (\sigma \cdot \ii \hat{\sigma} )(\sigma \cdot \sigma )
\right]
\right\} 
  \ ,
\eeqn
where $(\sigma \cdot \sigma ) \equiv \sum_{i=1}^N \sigma_i(t) \sigma_i(t')$, 
\beq
\alpha = \left\{
\begin{array}{ll}
\frac{1-pa^2}{1+a^2} & \, {\rm if }\ p\ {\rm is \ even}\ ,
\\
\frac{1-(p-1)a^2}{1+a^2} & \, {\rm if }\ p\ {\rm is \ odd}\ ,
\end{array}
\right.
\ ,
\eeq
and $\vec{n}_p$ denotes any $p$ non-repeating integers between 1 and $N$ in the ascending order.

Denoting Eq.~(\ref{eq:S_int}) by $\tri (\hat{\sigma}, \sigma)$, the overlap matrix $Q$'s are now introduced as follows:
\beqn
\tri (\hat{\sigma}, \sigma)
&=&
\int \cD Q \delta \left[NQ_1 -(\ii \hat{\sigma} \cdot  \ii \hat{\sigma})\right]
\delta \left[NQ_2 -({\sigma} \cdot  {\sigma})\right]
\delta \left[NQ_3 -(\ii \hat{\sigma} \cdot  {\sigma})\right]
\delta \left[NQ_4 -({\sigma} \cdot  \ii \hat{\sigma})\right]
\\
&&\times \exp \left\{ \frac{N}{4} \int \dd t \dd t' \left[p Q_1Q_2^{p-1} +p(p-1)\alpha Q_3 Q_4Q_2^{p-2}
\right]\right\}\ .
\eeqn
Now \cite{s_kirkpatrick_prl87,s_kirkpatrick_prb87},
\beq
Q_1(t,t')=0 \sep Q_2(t,t') =C(t,t') \sep Q_3(t,t')=R(t',t) \sep Q_4 (t,t')=R(t,t')
\eeq
Using further the exponential representation of the $\delta$-functions, e.g., 
\beq
\delta \left[NQ_2 -({\sigma} \cdot  {\sigma})\right]= \int D \lambda_2 \exp \left[ \ii N \int \dd t \dd t' \lambda_2 (Q_2-\sigma \cdot \sigma)\right]
\ .
\eeq
Saddle point method then leads to
\beqn
\ii \lambda_1 &=& \frac{p}{4} Q_2^2
\\
\ii \lambda_2 &=& \frac{p(p-1)}{4} \left[ Q_1 Q_2^{p-2} + (p-1)\alpha Q_3 Q_4 Q_2^{p-3}\right]
\\
\ii \lambda_3 &=& \frac{p(p-1)}{4} \alpha Q_4 Q_2^{p-2} 
\\
\ii \lambda_4 &=& \frac{p(p-1)}{4} \alpha Q_3 Q_2^{p-2} 
\eeqn
As in the thermal case \cite{s_kirkpatrick_prl87,s_kirkpatrick_prb87}, $Q_1=0, Q_3Q_4=0$, we thus arrive at the following dynamic equation
\beq
\pp_t \sigma=-\mu \sigma + \frac{\alpha p(p-1)}{2}\int \dd t' R(t,t')C(t,t')^{p-1}\sigma(t')+\xi
\eeq
where 
\beq
\la \xi(t)\xi(t')\ra =2 T \delta (t-t') + \frac{p}{2} C(t,t')^{p-1}
\eeq

From this we can obtain the dynamical equations for $C$ and $R$:
\beqn
\frac{\pp C(t,t')}{\pp t} &=& -\mu(t) C(t,t') + \frac{\alpha p(p-1)}{2}\int_{-\infty}^t  \dd t'' R(t,t'')C(t,t'')^{p-2}C(t'',t')
\\
&& +2TR(t',t)+\frac{p}{2} \int_{-\infty}^{t'}  \dd t'' R(t',t'')C(t,t'')^{p-1}
\\
\frac{\pp R(t,t')}{\pp t} &=& -\mu(t) R(t,t') + \frac{\alpha p(p-1)}{2}\int_{t'}^t  \dd t'' R(t,t'')C(t,t'')^{p-2}R(t'',t') +\delta(t,t')
\\
\mu(t) &=& T+\frac{\alpha p(p-1) +p}{2} \int_{-\infty}^{t}  \dd t'' R(t,t'')C(t,t'')^{p-1}
\eeqn

\section{Decay regimes of correlation function}
As in thermalizing supercooled fluids, numerical integration of the EOM (Fig.~2 in the MT) suggests that as the noise strength $T$ is lowered, the system's correlation $C(t, t')$ develops a plateau as $t$ evolves, and the plateau's duration lengthens as $t'$, the {\it waiting time}, increases. Furthermore, during the initial decay of $C$ to its plateau value, the fluctuation-dissipation relation is seen to be satisfied to a good degree, at least for small $a$ (i.e., $\alpha$ close to 1.) As a result, we will adopt the following approximations under the {\it weak ergodicity breaking scenario}, which are typically used in the study of thermalizing $p$-spin models \cite{s_bouchaud_jdp92,s_cugliandolo_prl93,s_cugliandolo_a02}:
 \beqn
 C(t,t')&=& C_{\rm eq}(t-t') + C_{\rm ag} (t,t')
 \\
 R(t,t')&=& R_{\rm eq}(t-t') + R_{\rm ag} (t,t')
 \eeqn 
where $C_{\rm eq}(\tau)$ and $R_{\rm eq}(\tau)$ account for the initial {\it equilibrium-like} dynamics, while $C_{\rm ag}(t,t')$ and $R_{\rm ag}(t,t')$ account for the dynamics in the {\it aging} regime. Specifically, we assume $C_{\rm eq}(0) = 1-q$, $R_{\rm eq} (\tau) = -\frac{1}{T} C_{\rm eq}'(\tau) \Theta(\tau)$, and $C_{\rm ag} (0,0)=q$, where $q$ is the plateau-height value of $C$, which is also termed the {\it Edward-Anderson} parameter.

\subsection{Glass transition}
The glass transition in the $p$-spin model is traditionally equated to the highest temperature $T$ below which $q$ first becomes nonzero.  

In the nonequilibrium model, to calculate this dynamical glass transition noise strength $T_d$, we can ignore the aging part of the correlation and response functions. From the EOM, we thus have for $t>t'$,
\beqn
\frac{\pp C_{\rm eq}(t-t')}{\pp t} &=& -\mu(t) C_{\rm eq}(t-t') - \frac{\alpha p(p-1)}{2T}\int_{-\infty}^t  \dd t'' C_{\rm eq}'(t-t'')C_{\rm eq}(t-t'')^{p-2}C_{\rm eq}(t''-t')
\\
&& -\frac{p}{2T} \int_{-\infty}^{t'}  \dd t'' C_{\rm eq}'(t'-t'')C_{\rm eq}(t-t'')^{p-1}
\\
\mu(t) &=& T-\frac{\alpha p(p-1) +p}{2T} \int_{-\infty}^{t}  \dd t''  C_{\rm eq}'(t-t'')C_{\rm eq}(t-t'')^{p-1}
\eeqn

Now for $\mu$,
\beq
\int_{-\infty}^{t}  \dd t''  C_{\rm eq}'(t-t'')C_{\rm eq}(t-t'')^{p-1}=-\frac{1}{p}\int_{-\infty}^{t} \dd  \left[ C_{\rm eq}(t-t'')^p\right]=
-\frac{1}{p} \left[C_{\rm eq}(0)^p - C_{\rm eq}(\infty)^p\right] = -\frac{1}{p} 
\ .
\eeq

Then, with $s=t''-t'$,
\beqn
\int_{-\infty}^t  \dd t'' C_{\rm eq}'(t-t'')C_{\rm eq}(t-t'')^{p-2}C_{\rm eq}(t''-t')
&=&-\frac{1}{p-1}\int_{-\infty}^\tau  \dd s  C_{\rm eq}(s)\frac{\dd }{\dd s} C_{\rm eq}(\tau-s)^{p-1} 
\\
&=&-\frac{1}{p-1}\int_{-\infty}^\tau  \dd s  C_{\rm eq}(s)\frac{\dd }{\dd s} C_{\rm eq}(\tau-s)^{p-1} 
\\
&=&-\frac{1}{p-1}\left[ C_{\rm eq}(\tau) -
\int_{-\infty}^\tau  \dd s C_{\rm eq}(\tau-s)^{p-1} \frac{\dd }{\dd s}  C_{\rm eq}(s) \right]
\ .
\eeqn

Finally, with $s=t'-t''$, we have
\beqn
\int_{-\infty}^{t'}  \dd t'' C_{\rm eq}'(t'-t'')C_{\rm eq}(t-t'')^{p-1}
&=&\int_{0}^\infty  \dd s  C_{\rm eq}(\tau+s)^{p-1}\frac{\dd }{\dd s} C_{\rm eq}(s)
\\
&=&-\int_{-\infty}^0  \dd s  C_{\rm eq}(\tau-s)^{p-1}\frac{\dd }{\dd s} C_{\rm eq}(s)
\ .
\eeqn

We thus have
\beqn
C_{\rm eq}'(\tau)&=& -\left(T +\frac{\alpha p(p-1) +p}{2Tp} \right) C_{\rm eq}(\tau) 
\\
&&+\frac{\alpha p}{2T} C_{\rm eq}(\tau) -\frac{\alpha p}{2T}  \int_{-\infty}^\tau\dd s C_{\rm eq}(\tau-s)^{p-1} C_{\rm eq}'(s)
\\
&&
 +\frac{ p}{2T}  \int_{-\infty}^0\dd s C_{\rm eq}(\tau-s)^{p-1} C_{\rm eq}'(s)
 \\
 &=& -\left(T +\frac{1-\alpha}{2T} \right)C_{\rm eq}(\tau) - \frac{p}{2T}  \int_{0}^\tau\dd s C_{\rm eq}(\tau-s)^{p-1} C_{\rm eq}'(s)
 +\frac{(1-\alpha) p}{2T}  \int_{-\infty}^\tau\dd s C_{\rm eq}(\tau-s)^{p-1} C_{\rm eq}'(s)
 \ .
\eeqn
If a plateau occurs for $C_{\rm eq}$, for large $\tau$, we expect the last integral to be zero since $C_{\rm eq}'(s)$ is symmetric.
In this regime, it is clear that the decay of $C_{\rm eq}$ is always faster whenever $a>0$. 

Using the approximation $\int_{A}^B \dd s C(\tau-s)^{p-1} C'(s)=C^{p-1}(\tau) \left[C(B)-C(A)\right]$, we have
\beq
-\left(T +\frac{1-a}{2T} \right)C(\tau) + \frac{p}{2T} C^{p-1}(\tau) \left[1-C(\tau)\right]
\leq 0\ .
 \eeq 
  Letting $q = \lim_{\tau \rightarrow \infty} C(\tau)$, we have equivalently the following
 \beq
 T^2 = \frac{(\alpha-1) +p(1-q)q^{p-2}}{2}
 \ .
 \eeq
The highest noise strength at which the $q$ admits a nonzero value occurs at $q=(p-2)/(p-1)$, and the corresponding threshold noise strength is
\beq
T_d = \sqrt{\frac{\alpha-1}{2} +\frac{p(p-2)^{p-2}}{2(p-1)^{p-1}}} \ .
\eeq

 \subsection{Aging regime}
 Below the critical noise strength $T_c$, we need to consider to consider the following approximate forms of the two-time correlation and response  functions:
 \beqn
 C(t,t')&=& C_{\rm eq}(t-t') + C_{\rm ag} (t,t')
 \\
 R(t,t')&=& R_{\rm eq}(t-t') + R_{\rm ag} (t,t')
 \eeqn
 such that $C_{\rm eq}(0) = 1-q$, $C_{\rm ag} (0,0)=q$, and $R_{\rm eq} (\tau) = -\frac{1}{T} C_{\rm eq}'(\tau)$.
 
 In the large $t$ and $t'$ limit in the stationary limit, we have using the EOM of the response function
 \beq 
 \label{eq:ag_r}
 \mu_\infty R_{\rm ag}(t,t) =
 \frac{\alpha p(p-1)}{2} \left[ q^{p-2} R_{\rm ag}(t,t) \int_0^\infty \dd \tau R_{\rm eq } (\tau) + R_{\rm ag}(t,t) \int_0^\infty \dd \tau R_{\rm eq} (\tau) C_{\rm eq}(\tau)^{p-2}\right]
 \eeq
or 
 \beq
 \label{eq:mu1}
\mu_\infty =
 \frac{\alpha p(p-1)}{2} \left[ q^{p-2}  \tau R_{\rm eq } (\omega=0) + \Sigma(\omega=0)\right]
 \eeq
 where
 \beqn
 R_{\rm eq } (\omega=0) &=& \int_0^\infty \dd \tau R_{\rm eq} (\tau) =\frac{1-q}{T}
 \\
 \Sigma(\omega=0) &=& \int_0^\infty \dd \tau R_{\rm eq} (\tau) C_{\rm eq}(\tau)^{p-2}=\frac{p(1-q)^{p-1}}{2T}
 \ .
 \eeqn
From the same EOM, we also have
\beq
\label{eq:mu2}
\mu_\infty R_{\rm eq}(\tau) =\delta(\tau)+ \frac{\alpha p(p-1)}{2} \int_0^\tau \dd \tau' R_{\rm eq} (\tau -\tau') C_{\rm eq} (\tau -\tau')^{p-2} R_{\rm eq} (\tau')
\ .
 \eeq
  Integrating $\tau$ in the above leads to
 \beq
 \mu_\infty R_{\rm eq}(\omega=0) = 1+\frac{\alpha p(p-1)}{2} \Sigma(\omega=0) R_{\rm eq} (\omega=0)
 \ .
 \eeq 
 Eqs.~(\ref{eq:mu1}) and (\ref{eq:mu2}) together give
 \beq
 T^2 = \frac{\alpha p(p-1)q^{p-2}(1-q)^2}{2}
 \ .
 \eeq
